\documentclass[a4paper]{article}
\usepackage{graphicx}
\usepackage{INTERSPEECH2020}
\usepackage{tabularx}
\usepackage{array}
\usepackage{longtable}
\usepackage{colortab}
\usepackage{colortbl}
\usepackage{arydshln}
\usepackage{multirow}
\usepackage{float}
\usepackage{bxbase}
\usepackage{flushend}

\title{Generative Adversarial Training Data Adaptation for Very Low-resource Automatic Speech Recognition}
\name{Kohei Matsuura, Masato Mimura, Shinsuke Sakai, Tatsuya Kawahara}
\address{Graduate School of Informatics, Kyoto University, Sakyo-ku, Kyoto, Japan}
\email{\{matsuura,mimura,sakai,kawahara\}@sap.ist.i.kyoto-u.ac.jp}

\begin{document}

\maketitle

\begin{abstract}
It is important to transcribe and archive speech data of endangered languages for preserving heritages of verbal culture and automatic speech recognition (ASR) is a powerful tool to facilitate this process.
However, since endangered languages do not generally have large corpora with many speakers, the performance of ASR models trained on them are considerably poor in general.
Nevertheless, we are often left with a lot of recordings of spontaneous speech data that have to be transcribed.
In this work, for mitigating this speaker sparsity problem, we propose to convert the whole training speech data and make it sound like the test speaker in order to develop a highly accurate ASR system for this speaker.
For this purpose, we utilize a CycleGAN-based non-parallel voice conversion technology to forge a labeled training data that is close to the test speaker's speech.
We evaluated this speaker adaptation approach on two low-resource corpora, namely, Ainu and Mboshi.
We obtained 35-60\% relative improvement in phone error rate on the Ainu corpus, and 40\% relative improvement was attained on the Mboshi corpus.
This approach outperformed two conventional methods namely unsupervised adaptation and multilingual training with these two corpora.
\end{abstract}

\section{Introduction}
It is important to transcribe and archive endangered languages for preserving the heritages of their verbal culture. Since it is highly expensive to manually transcribe a large amount of speech data of unfamiliar languages, there is a strong demand for an automatic speech recognition (ASR) system to facilitate the transcription process.
ASR performance, however, strongly depends on the training data, and a speech corpus of an endangered language generally has only a small number of speakers since there are not so many people who can speak it. As a result, when trained with such a small corpus, the ASR model cannot be generalized well and becomes poor at recognizing the speech of unknown speakers \cite{matsuura2020speech}.

In this work, we tackle the challenge of a typical problematic situation in very low-resource languages: there are transcribed speech data from only a few speakers and we have a new speaker whose oral recordings need to be transcribed. In order to handle this, we propose an effective speaker adaptation method which employs non-parallel voice conversion (VC) based on CycleGAN \cite{radford2015unsupervised, Zhu_2017_no_url}. The proposed approach consists
of two steps: (1) utterances in the training data are transformed to sound like the test speaker's voice, (2) the ASR model is trained
using the original and transformed data. Through these steps, the ASR model can learn an ``unknown'' speaker's voice in advance although it is an artificial one. For step (1), non-parallel VC is adopted since it does not require any parallel speech data, which is hard to construct with low-resource languages. Moreover, we use only a small part of the target speaker's speech in this step and the target speech in the test set remains untouched. As described, this method is a label-free, data-efficient, and completely unsupervised way of speaker adaptation. 
This is the first study to apply non-parallel VC-based speaker adaptation to real low-resource corpora.
We evaluated this method with the Ainu speech corpus \cite{matsuura2020speech} and the Mboshi corpus \cite{RIALLAND18.635_short}. 
Furthermore, we investigate how much target speaker's speech is needed for our VC-based approach to work effectively. 


\section{Speaker sparsity problem}

\subsection{Ainu speech corpus}
\begin{table}[t]\vspace{2pt}
\centering
\caption{Speaker-wise data distribution in Ainu corpus\vspace{-5pt}}
\begingroup
\setlength{\tabcolsep}{4.5pt}
\renewcommand{\arraystretch}{1.0} 
\begin{tabular}{lcccccccc} \hline\hline
Speaker ID & K& S$_1$& S$_2$& S$_3$ &S$_4$ &S$_5$ &U$_1$&U$_2$\\ \hline 
duration (h)  & 19.7&7.3&3.3&2.1&1.8&1.7&1.6&1.6 \\
duration (\%)  & 50.5&18.6&8.3&5.4&4.5&4.5&4.1&4.1 \\ \hline
\end{tabular}\vspace{-15pt}
\endgroup
\end{table}
The Ainu speech corpus \cite{matsuura2020speech} is a low-resource data set of an endangered language. It has only 8 speakers and the amount of the recordings is not balanced among speakers; instead more than half of the data is from only a single speaker (labeled ``K'') as seen in Table 1. In our previous work \cite{matsuura2020speech}, we evaluated the ASR performance on this corpus. We found that with the best modeling unit the performance was fairly good considering the limited amount of training data when speakers in the test set are included in the training set (the speaker-closed setting). However, when the test speakers were not included in the training set (the speaker-open setting), the recognition accuracy was significantly degraded due to the highly limited number of training speakers. In this paper, we work on this speaker sparsity problem, considering the situation where there are other Ainu speakers whose oral recordings are waiting to be transcribed.

Other endangered languages also do not have speech corpora with sufficient numbers of speakers. In our best knowledge of published speech corpora of endangered languages, the Griko corpus \cite{Boito2018ASG} has 9 speakers, the Mboshi corpus \cite{RIALLAND18.635_short} has 3 speakers, and the Basaa corpus \cite{HAMLAOUI18.948_short} is said to have ``a few speakers''. Therefore, the speaker sparsity problem widely appears in ASR of endangered languages besides Ainu.

\subsection{Conventional approaches}
We review speaker adaptation and multilingual training as conventional approaches to solve the speaker sparsity problem in sequence-to-sequence ASR.

In a widely adopted approach, a speaker-independent model is finetuned on the test data from new speakers using initial recognition results as labels. We refer to this method as \emph{self-supervised adaptation} in this paper.
Ochiai \textit{et al.\!\!} investigated which part of their combined speech enhancement and ASR model should be fixed considering the risk of overtraining \cite{8462161}. Meng \textit{et al.\!\!} introduced Kullback-Leibler divergence (KLD) regularization, adversarial speaker adaptation (ASA), and multi-task learning speaker adaptation to mitigate the overfitting \cite{Meng_2019}. In spite of these efforts, the ASR model is often affected by errors in the first-pass recognition results used as labels for adaptation data. 
Another popular way for speaker adaptation is appending i-vectors to input acoustic features \cite{Audhkhasi_2018}. 
The i-vector represents the specific characteristics of a speaker's voice, and it is calculated with the universal background model (UBM), which generally requires many speakers and is difficult to construct in low-resource languages. 
Other feature-space adaptation methods such as maximum likelihood linear regression \cite{6af3452a28a04980b2b8f5eb48730d36} and maximum a posteriori adaptation using GMM-derived features \cite{Tomashenko2014SpeakerAO, tomashenko-esteve-2018-evaluation_short} are not as effective as model retraining in a low-resource situation.

It is well-known that the performance of low-resource ASR is improved by using corpora of other languages. This method is called \emph{multilingual training}. Typically, one ASR model is shared with multiple languages and it picks up an output label from the union of grapheme sets of the languages \cite{Kim_2018_short, Toshniwal_2018_short, 8268945}.
We examined the effectiveness of multilingual training on the Ainu corpus and obtained some improvement \cite{matsuura2020speech}.

\section{Non-parallel voice conversion approach}

\subsection{Basic concept and processing flow}
As mentioned in the previous section, self-supervised adaptation exploits the matched data but easily overfits them and is also error-prone. On the other hand, multilingual training does not augment data of the very target language for training.
To overcome the drawbacks of these two approaches, we adopt an approach of unsupervised adaptation to generate data matched to test speakers without relying on erroneous labels or using a large number of speakers, or data of other languages. Instead, the proposed approach attempts to convert the existing data of one or a few speakers in the training set to the new speaker in the test set. 
This idea is simple but has not been practical when the quality of voice conversion is not good.

In recent years, high-quality non-parallel VC methods based on CycleGAN \cite{radford2015unsupervised,Zhu_2017_no_url} have been introduced \cite{8553236, kaneko2019CycleGAN-VC2}. They do not require parallel utterance pairs, which are generally not available in low-resource languages.
Therefore, we investigate the CycleGAN-based approach in very low-resource situations.

The procedure of the proposed method is as follows. First, we prepare source and target acoustic features ($S$ and $T$, respectively) to train CycleGAN. $S$ is from original training data, and $T$ is from the target speaker who is in the test set and unseen in the training set. 
The CycleGAN is trained to minimize the loss described in the next section to obtain a generator with which utterances in $S$ are transformed to have characteristics of utterances in $T$.
After the training of the CycleGAN, all features in the training data are converted using the generator. Finally, the ASR model is trained with converted and original training data.


\subsection{CycleGAN-based non-parallel voice conversion}
\begin{figure}[t]
  \centering
  \includegraphics[width=0.9\linewidth]{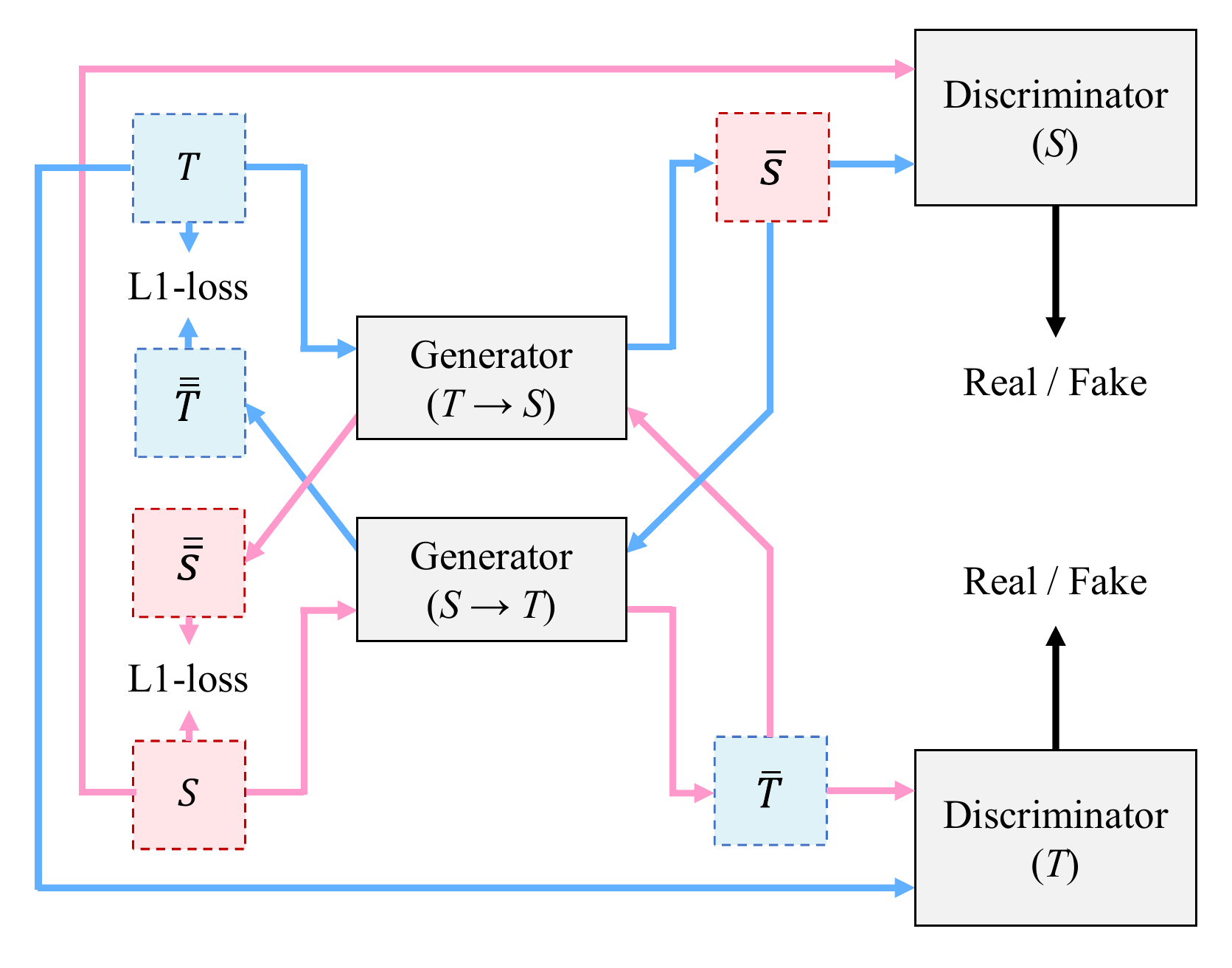}\vspace{-8pt}
  \caption{The architecture of CycleGAN. $S$ and $T$ are the source and target speaker's features, respectively. $\overline{X}$ denotes generated fake features of $X$. The red paths start from $S$ and the blue ones from $T$. The identity-mapping loss is not described here.\vspace{-12pt}}
  \label{fig:prop}
\end{figure}
In this section, we explain the details of CycleGAN. 
CycleGAN has two generators and two discriminators. Generators convert source/target speaker's voice into the target/source speaker's voice, and discriminators judge whether the input voice is from a real dataset or a generator as shown in Figure 1. 

In the following equations, $S$ and $T$ represent the total sets of source and target speaker's features, respectively. $\mathrm{G}_{S \rightarrow T}$ means the generator which is trained to convert source speaker's voice into target speaker's voice. $\mathrm{D}_S$ is the discriminator for real and fake source speaker's voices. Note that 1 and 0 are labels for real data and fake data in equations (1) and (5).
Generators are trained with the following three objectives:
\begin{enumerate}
\item{\textbf{Adversarial Loss}: This loss encourages generators to output more confusing features for discriminators. The least mean square error is used following LSGAN \cite{Mao_2017}.}
\begin{eqnarray}
\mathcal{L}_{\mathrm{G} (\mathrm{adv})} \!\!\!\!&=&\!\!\!\! \mathop{\mathbb{E}}_{s \sim p_S (s)}[(\mathrm{D}_T (\mathrm{G}_{S \rightarrow T} (s)) - 1)^2] \nonumber \\
\!\!&+&\!\!\!\! \mathop{\mathbb{E}}_{t \sim p_T (t)}[(\mathrm{D}_S (\mathrm{G}_{T \rightarrow S} (t)) - 1)^2]
\end{eqnarray}
\item{\textbf{Cycle-consistency Loss}: With this loss, input features can be reconstructed after passing through two generators. The linguistic consistency between generators' input and output is expected to be maintained.}
\begin{eqnarray}
\mathcal{L}_{\mathrm{G} (\mathrm{cyc})} \!\!\!\!&=&\!\!\!\! \mathop{\mathbb{E}}_{s \sim p_{S} (s)}[|| \mathrm{G}_{T \rightarrow S} (\mathrm{G}_{S \rightarrow T} (s)) - s||_1] \nonumber \\
\!\!\!\!&+&\!\!\!\! \mathop{\mathbb{E}}_{t \sim p_T (t)}[|| \mathrm{G}_{S \rightarrow T} (\mathrm{G}_{T \rightarrow S} (t)) - t||_1]
\end{eqnarray}
\item{\textbf{Identity-mapping Loss}: This loss requires generators to avoid unnecessary modification of input features.}
\begin{eqnarray}
\mathcal{L}_{\mathrm{G} (\mathrm{id})} \!\!\!\!&=&\!\!\!\! \mathop{\mathbb{E}}_{t \sim p_T (t)}[|| \mathrm{G}_{S \rightarrow T} (t) - t ||_1] \nonumber \\
\!\!&+&\!\!\!\! \mathop{\mathbb{E}}_{s \sim p_S (s)}[|| \mathrm{G}_{T \rightarrow S} (s) - s ||_1]
\end{eqnarray}
\end{enumerate}
With hyperparameters $\lambda_\mathrm{cyc}$ and $\lambda_\mathrm{id}$, the entire loss of the generators is defined as below:
\begin{eqnarray}
\mathcal{L}_{\mathrm{G}} = \mathcal{L}_{\mathrm{G}(\mathrm{adv})} + \lambda_\mathrm{cyc}\mathcal{L}_{\mathrm{G}(\mathrm{cyc})} + \lambda_\mathrm{id}\mathcal{L}_{\mathrm{G\mathrm{(id)}}}
\end{eqnarray}
Discriminators are trained with the following loss:
\begin{eqnarray}
\mathcal{L}_{\mathrm{D}} \!\!\!\!&=&\!\!\!\!\!\! \mathop{\mathbb{E}}_{s \sim p_S (s)}[(\mathrm{D}_{T} (\mathrm{G}_{S \rightarrow T} (s)) - 0)^2 + (\mathrm{D}_{S} (s) - 1)^2]\nonumber\\
\!\!\!\!&+&\!\!\!\!\!\! \mathop{\mathbb{E}}_{t \sim p_T (t)}[(\mathrm{D}_{S} (\mathrm{G}_{T \rightarrow S} (t)) - 0)^2 + (\mathrm{D}_{T} (t) - 1)^2]
\end{eqnarray}
When generators' parameters are updated with $\mathcal{L}_{\mathrm{G}}$, discriminators' parameters are left unchanged, and vice versa. 

For generators and discriminators, we employ CycleGAN-VC2 \cite{kaneko2019CycleGAN-VC2}, which is a CycleGAN specially developed for voice conversion, unlike the networks adopted in related works in Section 3.3. Specifically, its generators have 2-1-2D CNN architecture, which is the combination of 1D CNN for modeling dynamic change in speech signals and 2D CNN for preserving the original structure. ParchGAN \cite{DBLP:journals/corr/LiW16b, DBLP:journals/corr/ShiCHTABRW16} is used for its discriminators. We do not use ``two-step adversarial loss" proposed in \cite{kaneko2019CycleGAN-VC2} because we found it not so helpful for ASR in a preliminary experiment.

\vspace{-3pt}
\subsection{Related works}
\vspace{-1.5pt}
There has been some previous work on adaptation with CycleGAN-based feature mapping.
Mimura \textit{et al.\!} adopted it for domain adaptation including speech enhancement. This was the first work to use CycleGAN for front-end feature transformation in ASR \cite{8268927}. Dumpara \textit{et al.} extracted
perturbed speech from the AMI meeting corpus and the Buckeye corpus and observed
that the degraded ASR performance with such speech was improved by a CycleGAN-based front-end \cite{dumpala2019cyclegan}.
Hosseini-Asl \textit{et al.} investigated the effect of the CycleGAN-based adaptation between two genders by manually separating the TIMIT corpus into male and female speakers \cite{hosseini-asl2019augmented}. 
Our work reported in this paper is the first study that applies a CycleGAN-based voice cloning technique to individual speaker adaptation for ASR and demonstrates its effectiveness for very low-resource ASR in a practical situation.
\vspace{-2.5pt}
\section{Experimental evaluations}
\subsection{Dataset}
\vspace{-1.5pt}
We evaluated the proposed VC-based approach through speech recognition experiments using the Ainu corpus and the Mboshi corpus. The Ainu corpus contains about 40 hours of folklore recited by 8 speakers. From among them, 2 different speakers were chosen as unknown target speakers (U$_i$, $i = \{1, 2\}$), and transcribed speech data from the other speakers (ALL/U$_i$) were used for the training set in the experiment. In addition, we tried an extreme situation where we assumed we had only one ``known'' (i.e. labeled) speaker (K), who has the largest amount of data among the Ainu speakers. We performed four experiments with the Ainu corpus (Table 2), where experimental IDs of K-U$_i$ and ALL-U$_i$ are given for convenience. CycleGAN was trained using all of the entire known speaker's features and randomly chosen 1/5/10/20/30 minutes of the target speaker's features that are set aside from the test set. 

Furthermore, the Mboshi corpus \cite{RIALLAND18.635_short} was chosen to see whether this approach is effective in other low-resource languages. The Mboshi corpus contains about 5 hours of speech read by 3 speakers that we refer to as A, B, and C. The data portion for each speaker is divided into training and development set in the corpus and we adopt this ``official'' definitions of subsets.
We designated the speaker C, who speaks the least, as the unknown test speaker and speaker A and B as known training speakers in the experiment. The CycleGAN was trained using the training set portion of the data to learn the conversion from training speakers A and B to test speaker C. The ASR model was trained using this converted data and evaluated using the development set of C.
\begin{table}[t]
\centering
\caption{Four experimental settings in the Ainu corpus \vspace{-5pt}}
\begingroup
\setlength{\tabcolsep}{7.0pt}
\renewcommand{\arraystretch}{1.1} 
\begin{tabular}{lcccc} \hline\hline
Experiment ID & K-$\mathrm{U}_1$  & K-$\mathrm{U}_2$   & ALL-U$_1$& ALL-U$_2$\\ \hline 
known spkr.  & K& K & ALL/U$_1$  & ALL/U$_2$ \\
~~\# speakers & 1&1&7&7\\ 
~~duration (h)  & 19.68  & 19.68 & 37.18 & 37.17  \\ \hdashline 
target spkr. &U$_1$   & U$_2$ & U$_1$ &U$_2$  \\ \hline
\end{tabular}
\endgroup\vspace{-13pt}
\end{table}
\vspace{-3.5pt}
\subsection{Experimental details}
\vspace{-1.5pt}
In CycleGAN training, acoustic features are 40-dimensional log Mel filter banks (MFBs) extracted every 10 ms over a 25-ms window. While generators convert an entire sequence, discriminators accept cropped 128 frames of features. In Eq. (4), $\lambda_\mathrm{id}$ is 5 only for the first $10^4$ iterations and then set to 0, while $\lambda_\mathrm{cyc}$ is 10 throughout the training. We trained the networks for $5\!\times\!10^{4}$ steps with the Adam optimizer \cite{Adam} with a batch size of 5. The learning rate for generators is $2\!\times\!10^{-4}$ and that for discriminators is $1\!\times\!10^{-4}$. The ASR model is an attention-based encoder-decoder model \cite{6af3452a28a04980b2b8f5eb48730d36, AttnProto, 7472621} with Connectionist Temporal Classification (CTC) \cite{GravesCTC, Graves2014TowardsES} subtasks \cite{DBLP:journals/corr/KimHW16}. 
In ASR training, we stack 3 consecutive input frames to form a sequence of 120-dimensional features \cite{frame}.
The encoder is a 5-layer bidirectional long short-term memory (LSTM) \cite{Hochreiter:1997:LSM:1246443.1246450, 650093} and the decoder is a 1-layer unidirectional LSTM. All LSTMs have 320 hidden units. The 1D convolution layer in the location-based attention mechanism has 10 channels and their kernel width is 100. Dropout \cite{dropout} of 0.2 is applied to the encoder LSTM. The total loss $\mathcal{L}$ is a weight sum of the main loss $\mathcal{L}_\mathrm{attn}$ and the loss of CTC subtask $\mathcal{L}_\mathrm{ctc}$:
\setlength{\abovedisplayskip}{5pt} 
\setlength{\belowdisplayskip}{5pt} 
\begin{eqnarray}
\mathcal{L} = 0.8\mathcal{L}_\mathrm{attn} + 0.2\mathcal{L}_\mathrm{ctc}\vspace{-5pt}
\end{eqnarray}
We chose syllable for the modeling unit following \cite{matsuura2020speech} on the Ainu corpus, and chose phone on the Mboshi corpus. The modeling unit for CTC subtasks is phone for the both corpora. 
For the baseline experiment, we evaluated the ASR model trained only with the original training data shown in Table 2. We trained the networks for 60 epochs with weight decay \cite{Krogh92asimple} of $1\!\times\!10^{-5}$. The learning rate is $1\!\times\!10^{-3}$ for the first 30 epochs and is then multiplied by 0.9 at the beginning of each epoch. 
In the proposed approach, the whole converted data are added to original training data for ASR training.
Self-supervised adaptation and multilingual training are compared with the proposed approach. In self-supervised adaptation, the baseline ASR model was finetuned with the same learning rate scheduling as mentioned above. KLD regularization and ASA \cite{Meng_2019} in Section 2.2 were not applied because we found in preliminary experiments that they were not so helpful on the Ainu and Mboshi corpora. In multilingual training, the English corpus WSJ \cite{wsj} and the Japanese corpus JNAS \cite{jnas} are used. JNAS comprises about 80 hours of speech from 324 speakers, and WSJ has about 70 hours of speech from 282 speakers. The encoder and the attention mechanism are shared with the three languages as in \cite{matsuura2020speech}. The modeling unit for English and Japanese is phones. 

\subsection{Results and discussions}
\vspace{-1.5pt}
First, we show the speech recognition result on the Ainu corpus in Table 3. The numbers for `self-supervised' and `VC' are the best PERs among 5 different amounts (i.e. 1/5/10/20/30 minutes) of data for adaptation. In all of the four experiments, the VC-based approach with CycleGAN yields drastically better results than other methods. In experiments K-U$_1$ and K-U$_2$, the PERs are improved from 45.4\% to 17.8\% (60.6\% relative improvement) and from 42.6\% to 18.3\% (57.0\% relative improvement), respectively, with 30 minutes of target speaker's data. In experiments ALL-U$_1$ and ALL-U$_2$, the baseline model performs much better than those in experiments K-U$_1$ and K-U$_2$. This means that the number of speakers in training data is critical for recognizing an unknown speaker's speech. Although the baseline model already performs well, our proposed method improves it further as in Table 3. For instance, the PER is decreased from 15.9\% to 10.5\% (33.7\% relative improvement) in experiment ALL-U$_1$ and from 13.8\% to 8.8\% (35.9\% relative improvement) in experiment ALL-U$_2$ with 20 minutes of target speaker's features. 
Note that this result is obtained without using the test sets in the VC training.
Therefore, when additional data from the same test speaker is found and to be recognized, there is no need to apply this VC-based adaptation again.
\begin{table}[t]
\centering
\caption{The PERs (\%) on the Ainu corpus. The numbers for `self-supervised' and `VC' are the best PERs among 5 different amounts (i.e. 1/5/10/20/30 minutes) of data for adaptation.\vspace{-3pt}}
\begingroup
\setlength{\tabcolsep}{6.1pt}
\renewcommand{\arraystretch}{1.15} 
\begin{tabular}{lcccc} \hline\hline
\multicolumn{1}{l}{Experiment ID} &K-$\mathrm{U}_1$  & K-$\mathrm{U}_2$   & ALL-U$_1$& ALL-U$_2$ \\ \hline 
baseline & 45.4 & 42.6 & 15.9 & 13.8 \\
self-supervised  & 45.6 & 38.8 & 14.7 & 12.0\\
multilingual  & 30.4 & 33.6 & 13.2 & 11.1 \\ 
VC &\textbf{17.8} &\textbf{18.3}&\textbf{10.5}&\textbf{8.8}\\ \hline
\end{tabular}
\endgroup
\end{table}
\begin{figure}[t]
  \centering
  \includegraphics[width=0.95\linewidth]{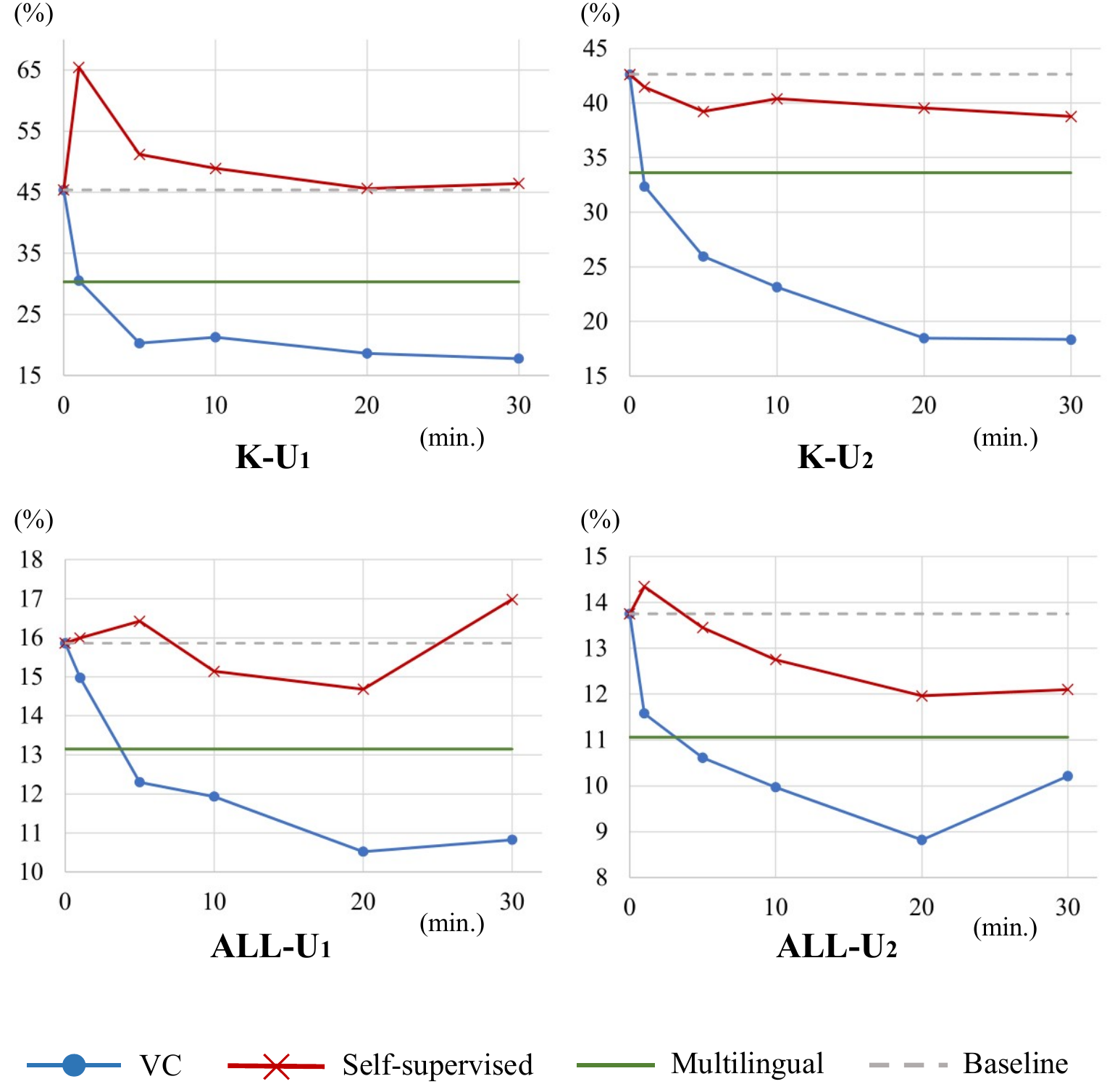}
  \caption{The PERs (\%) of the VC, self-supervised, and multilingual approaches with the baseline in four experiments on the Ainu corpus. VC and self-supervised adaptation have multiple results with 1/5/10/20/30-minute target speaker's features.}
  \label{fig:prop}\vspace{-10pt}
\end{figure}

The PERs with various lengths of target speaker's speech for the VC training and self-supervised adaptation are shown in Figure 2. In all experiments, 20 minutes of the target speaker's features look enough for the convergence of performance. This demonstrates the data-efficient nature of the proposed method and suggests that the CycleGAN VC-based approach can be applied to a wide range of practical low-resource situations. The self-supervised adaptation is not very effective in low-resource ASR. This is probably because the self-supervised adaptation requires a first-pass decoding result for adaptation data with a certain level of accuracy, which is rarely reached with low-resource datasets. 
In all settings, the VC-based method with only 5-minute adaptation data from a target speaker outperforms the multilingual training.

We show an example of voice conversion in the 40-dimensional log MFB domain in Figure 3. Here, (a) is speaker K's original speech and (b) is speaker U$_1$-like converted speech. While the original speech tends to have high energies around the middle-frequency bins, the converted speech does not have such a trend as seen in the red dashed squares in Figure 3. 
In Table 4, we show an example of improvement seen in experiment $\mathrm{K}$-$\mathrm{U}_1$. This sentence is the first two utterances in the test set of speaker U$_1$. 
Despite some errors, deletions are significantly decreased and the results are much more useful for making a transcript.

Table 5 shows the result on the Mboshi corpus. The proposed CycleGAN VC-based approach improves the PER by relatively 41.1\% compared with the baseline, and it has a trend similar to that of the results on the Ainu corpus. This demonstrates that the effectiveness of the CycleGAN VC-based approach in the very low-resource situation is not limited to a specific language. This experiment can be reproduced with our model and recipe located here\footnote{https://github.com/Kohei-Matsuura/Non-parallel-VC-on-Mboshi}.
\begin{figure}[t]
  \centering
  \includegraphics[width=0.95\linewidth]{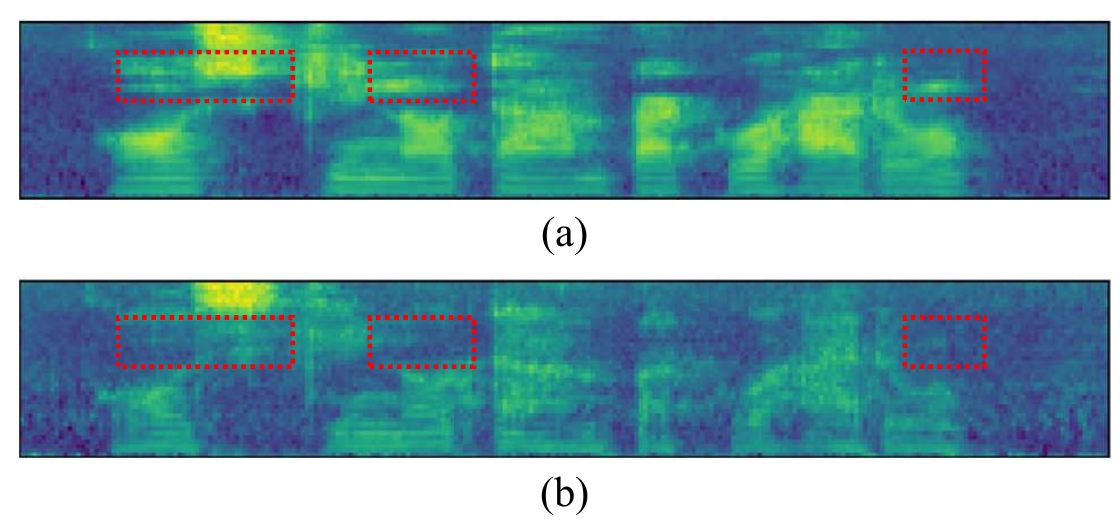}\vspace{-6pt}
  \caption{The comparison between speaker K's original feature (a) and speaker U$_1$-like fake feature (b). In the red dashed squares, (b) has lower energies than (a).\vspace{-2pt}}
  \label{fig:prop}
\end{figure}
\begin{table}[t]
\centering
\caption{An example of improvement. The VC result has far fewer deletion errors than the baseline result. \vspace{-5pt}}
\begingroup
\setlength{\tabcolsep}{5pt}
\renewcommand{\arraystretch}{1.2} 
\begin{tabular}{|ll|} \hline
ground truth & \scriptsize \it{a unuhu an a onaha an hine oka an hike iskar emko un} \\
baseline &  \scriptsize \it{a onaha ne okkaymi ki iskar emko}\\
VC & \scriptsize \it{a ponomo an a onaha an hine oka an he ki iskar emko un}\\ \hline
\end{tabular}
\endgroup
\end{table}
\begin{table}[t!]
\centering
\caption{The PERs (\%) on the Mboshi corpus \vspace{-5pt}}
\begingroup
\setlength{\tabcolsep}{7.5pt}
\renewcommand{\arraystretch}{1.13} 
\begin{tabular}{lc} \hline\hline
baseline & 44.0 \\
self-supervised  & 43.3\\
multilingual  & 34.6  \\ 
VC &\textbf{25.9}\\ \hline
\end{tabular}\vspace{-10pt}
\endgroup
\end{table}

\section{Conclusion}
In this work, we proposed a non-parallel VC-based approach with CycleGAN for speaker adaptation in the situation where there are only a very limited number of speakers in the corpus.
In this adaptation method, acoustic features in the training data are converted to target speaker-like data via the generator of CycleGAN, and then the ASR model is trained with the original and the converted training data. Comparing with conventional self-supervised adaptation and multilingual training, we demonstrated that the proposed approach is the most effective among these to mitigate the speaker sparsity problem on the Ainu corpus. 
This approach brings significant improvement from the baseline to the level to be used for transcriptions. 
In addition, we observed the same trend of results with the Mboshi corpus. This suggests that non-parallel VC-based speaker adaptation will be effective in ASR of various endangered languages. 

\newpage
\bibliographystyle{IEEEtran}

\bibliography{my}

\begin{thebibliography}{10}
\providecommand{\url}[1]{#1}
\csname url@samestyle\endcsname
\providecommand{\newblock}{\relax}
\providecommand{\bibinfo}[2]{#2}
\providecommand{\BIBentrySTDinterwordspacing}{\spaceskip=0pt\relax}
\providecommand{\BIBentryALTinterwordstretchfactor}{4}
\providecommand{\BIBentryALTinterwordspacing}{\spaceskip=\fontdimen2\font plus
\BIBentryALTinterwordstretchfactor\fontdimen3\font minus
  \fontdimen4\font\relax}
\providecommand{\BIBforeignlanguage}[2]{{%
\expandafter\ifx\csname l@#1\endcsname\relax
\typeout{** WARNING: IEEEtran.bst: No hyphenation pattern has been}%
\typeout{** loaded for the language `#1'. Using the pattern for}%
\typeout{** the default language instead.}%
\else
\language=\csname l@#1\endcsname
\fi
#2}}
\providecommand{\BIBdecl}{\relax}
\BIBdecl

\bibitem{matsuura2020speech}
K.~Matsuura, S.~Ueno, M.~Mimura, S.~Sakai, and T.~Kawahara, ``Speech corpus of
  {A}inu folklore and end-to-end speech recognition for {A}inu language,'' in
  \emph{Proceedings of The 12th Language Resources and Evaluation
  Conference}.\hskip 1em plus 0.5em minus 0.4em\relax European Language
  Resources Association, May 2020, pp. 2615--2621.

\bibitem{radford2015unsupervised}
A.~Radford, L.~Metz, and S.~Chintala, ``Unsupervised representation learning
  with deep convolutional generative adversarial networks,'' 2015,
  arxiv:1511.06434.

\bibitem{Zhu_2017_no_url}
J.-Y. Zhu, T.~Park, P.~Isola, and A.~A. Efros, ``Unpaired image-to-image
  translation using cycle-consistent adversarial networks,'' \emph{2017 IEEE
  International Conference on Computer Vision (ICCV)}, Oct 2017.

\bibitem{RIALLAND18.635_short}
A.~Rialland, M.~Adda-Decker, G.-N. Kouarata, G.~Adda, L.~Besacier, L.~Lamel,
  E.~Gauthier, P.~Godard, and J.~Cooper-Leavitt, ``{Parallel Corpora in Mboshi
  (Bantu C25, Congo-Brazzaville)},'' in \emph{LREC}, Miyazaki, Japan, May 7-12
  2018.

\bibitem{Boito2018ASG}
M.~Z. Boito, A.~Anastasopoulos, M.~Lekakou, A.~Villavicencio, and L.~Besacier,
  ``A small {G}riko-{I}talian speech translation corpus,'' in \emph{SLTU},
  2018.

\bibitem{HAMLAOUI18.948_short}
F.~Hamlaoui, E.-M. Makasso, M.~M\"{u}ller, J.~Engelmann, G.~Adda, A.~Waibel,
  and S.~St\"{u}ker, ``{BULBasaa: A Bilingual Basaa-French Speech Corpus for
  the Evaluation of Language Documentation Tools},'' in \emph{LREC}, Miyazaki,
  Japan, May 7-12 2018.

\bibitem{8462161}
T.~{Ochiai}, S.~{Watanabe}, S.~{Katagiri}, T.~{Hori}, and J.~{Hershey},
  ``Speaker adaptation for multichannel end-to-end speech recognition,'' in
  \emph{2018 IEEE International Conference on Acoustics, Speech and Signal
  Processing (ICASSP)}, 2018, pp. 6707--6711.

\bibitem{Meng_2019}
Z.~Meng, Y.~Gaur, J.~Li, and Y.~Gong, ``Speaker adaptation for attention-based
  end-to-end speech recognition,'' \emph{Interspeech 2019}, Sep 2019.

\bibitem{Audhkhasi_2018}
K.~Audhkhasi, B.~Kingsbury, B.~Ramabhadran, G.~Saon, and M.~Picheny, ``Building
  competitive direct acoustics-to-word models for english conversational speech
  recognition,'' \emph{2018 IEEE International Conference on Acoustics, Speech
  and Signal Processing (ICASSP)}, Apr 2018.

\bibitem{6af3452a28a04980b2b8f5eb48730d36}
J.~Chorowski, D.~Bahdanau, K.~Cho, and Y.~Bengio, ``\BIBforeignlanguage{English
  (US)}{End-to-end continuous speech recognition using attention-based
  recurrent {NN}: First results},'' in \emph{\BIBforeignlanguage{English
  (US)}{NIPS 2014 Workshop on Deep Learning}}, 2014.

\bibitem{Tomashenko2014SpeakerAO}
N.~A. Tomashenko and Y.~Y. Khokhlov, ``Speaker adaptation of context dependent
  deep neural networks based on map-adaptation and {GMM}-derived feature
  processing,'' in \emph{INTERSPEECH}, 2014.

\bibitem{tomashenko-esteve-2018-evaluation_short}
N.~Tomashenko and Y.~Est{\`e}ve, ``Evaluation of feature-space speaker
  adaptation for end-to-end acoustic models,'' in \emph{LREC}, Miyazaki, Japan,
  may 2018.

\bibitem{Kim_2018_short}
S.~Kim and M.~L. Seltzer, ``Towards language-universal end-to-end speech
  recognition,'' \emph{ICASSP}, Apr 2018.

\bibitem{Toshniwal_2018_short}
S.~Toshniwal, T.~N. Sainath, R.~J. Weiss, B.~Li, P.~Moreno, E.~Weinstein, and
  K.~Rao, ``Multilingual speech recognition with a single end-to-end model,''
  \emph{ICASSP}, Apr 2018.

\bibitem{8268945}
S.~{Watanabe}, T.~{Hori}, and J.~R. {Hershey}, ``Language independent
  end-to-end architecture for joint language identification and speech
  recognition,'' in \emph{2017 IEEE Automatic Speech Recognition and
  Understanding Workshop (ASRU)}, 2017, pp. 265--271.

\bibitem{8553236}
T.~{Kaneko} and H.~{Kameoka}, ``Cyclegan-vc: Non-parallel voice conversion
  using cycle-consistent adversarial networks,'' in \emph{2018 26th European
  Signal Processing Conference (EUSIPCO)}, 2018, pp. 2100--2104.

\bibitem{kaneko2019CycleGAN-VC2}
T.~Kaneko, H.~Kameoka, K.~Tanaka, and N.~Hojo, ``Cycle{GAN}-{VC}2: Improved
  cyclegan-based non-parallel voice conversion,'' in \emph{Proceedings of the
  IEEE International Conference on Acoustics, Speech and Signal Processing},
  2019.

\bibitem{Mao_2017}
X.~Mao, Q.~Li, H.~Xie, R.~Y. Lau, Z.~Wang, and S.~P. Smolley, ``Least squares
  generative adversarial networks,'' \emph{2017 IEEE International Conference
  on Computer Vision (ICCV)}, Oct 2017.

\bibitem{DBLP:journals/corr/LiW16b}
C.~Li and M.~Wand, ``Precomputed real-time texture synthesis with {M}arkovian
  generative adversarial networks,'' \emph{CoRR}, vol. abs/1604.04382, 2016.

\bibitem{DBLP:journals/corr/ShiCHTABRW16}
W.~Shi, J.~Caballero, F.~Husz{\'{a}}r, J.~Totz, A.~P. Aitken, R.~Bishop,
  D.~Rueckert, and Z.~Wang, ``Real-time single image and video super-resolution
  using an efficient sub-pixel convolutional neural network,'' \emph{CoRR},
  vol. abs/1609.05158, 2016.

\bibitem{8268927}
M.~{Mimura}, S.~{Sakai}, and T.~{Kawahara}, ``Cross-domain speech recognition
  using nonparallel corpora with cycle-consistent adversarial networks,'' in
  \emph{2017 IEEE Automatic Speech Recognition and Understanding Workshop
  (ASRU)}, 2017, pp. 134--140.

\bibitem{dumpala2019cyclegan}
S.~H. Dumpala, I.~Sheikh, R.~Chakraborty, and S.~K. Kopparapu,
  ``Cycle-consistent gan front-end to improve asr robustness to perturbed
  speech,'' in \emph{32nd Conference on Neural Information Processing Systems
  (NIPS)}, Montr\'{e}al, Canada, 2018.

\bibitem{hosseini-asl2019augmented}
E.~Hosseini-Asl, Y.~Zhou, C.~Xiong, and R.~Socher, ``Augmented cyclic
  adversarial learning for low resource domain adaptation,'' \emph{ICLR}, 2019.

\bibitem{Adam}
D.~P. Kingma and J.~Ba, ``Adam: {A} method for stochastic optimization,''
  \emph{CoRR}, vol. abs/1412.6980, 2014.

\bibitem{AttnProto}
D.~Bahdanau, J.~Chorowski, D.~Serdyuk, P.~Brakel, and Y.~Bengio, ``End-to-end
  attention-based large vocabulary speech recognition,'' in \emph{2016 IEEE
  International Conference on Acoustics, Speech and Signal Processing
  (ICASSP)}, March 2016, pp. 4945--4949.

\bibitem{7472621}
W.~{Chan}, N.~{Jaitly}, Q.~{Le}, and O.~{Vinyals}, ``Listen, attend and spell:
  A neural network for large vocabulary conversational speech recognition,'' in
  \emph{2016 IEEE International Conference on Acoustics, Speech and Signal
  Processing (ICASSP)}, 2016, pp. 4960--4964.

\bibitem{GravesCTC}
A.~Graves, S.~Fern{\'a}ndez, F.~J. Gomez, and J.~Schmidhuber, ``Connectionist
  temporal classification: labelling unsegmented sequence data with recurrent
  neural networks,'' in \emph{ICML}, 2006.

\bibitem{Graves2014TowardsES}
A.~Graves and N.~Jaitly, ``Towards end-to-end speech recognition with recurrent
  neural networks,'' in \emph{ICML}, 2014.

\bibitem{DBLP:journals/corr/KimHW16}
S.~Kim, T.~Hori, and S.~Watanabe, ``Joint ctc-attention based end-to-end speech
  recognition using multi-task learning,'' \emph{CoRR}, vol. abs/1609.06773,
  2016.

\bibitem{frame}
X.~Tian, J.~Zhang, Z.~Ma, Y.~He, and J.~Wei, ``Frame stacking and retaining for
  recurrent neural network acoustic model,'' \emph{CoRR}, vol. abs/1705.05992,
  2017.

\bibitem{Hochreiter:1997:LSM:1246443.1246450}
S.~Hochreiter and J.~Schmidhuber, ``Long short-term memory,'' \emph{Neural
  Comput.}, vol.~9, no.~8, pp. 1735--1780, Nov. 1997.

\bibitem{650093}
M.~{Schuster} and K.~K. {Paliwal}, ``Bidirectional recurrent neural networks,''
  \emph{IEEE Transactions on Signal Processing}, vol.~45, no.~11, pp.
  2673--2681, 1997.

\bibitem{dropout}
N.~Srivastava, G.~Hinton, A.~Krizhevsky, I.~Sutskever, and R.~Salakhutdinov,
  ``Dropout: A simple way to prevent neural networks from overfitting,''
  \emph{Journal of Machine Learning Research}, vol.~15, pp. 1929--1958, 2014.

\bibitem{Krogh92asimple}
A.~Krogh and J.~A. Hertz, ``A simple weight decay can improve generalization,''
  in \emph{ADVANCES IN NEURAL INFORMATION PROCESSING SYSTEMS 4}.\hskip 1em plus
  0.5em minus 0.4em\relax Morgan Kaufmann, 1992, pp. 950--957.

\bibitem{wsj}
D.~B. Paul and J.~M. Baker, ``The design for the wall street journal-based
  {CSR} corpus,'' in \emph{Speech and Natural Language: Proceedings of a
  Workshop Held at Harriman}, 1992.

\bibitem{jnas}
K.~Itou, M.~Yamamoto, K.~Takeda, T.~Takezawa, T.~Matsuoka, T.~Kobayashi,
  K.~Shikano, and S.~Itahashi, ``{JNAS}: {J}apanese speech corpus for large
  vocabulary continuous speech recognition research,'' \emph{Journal of the
  Acoustical Society of Japan (E)}, vol.~20, no.~3, pp. 199--206, 1999.

\end{thebibliography}

\end{document}